\documentclass[fleqn,12pt,twoside]{article}
\usepackage[headings]{espcrc1}

\readRCS
$Id: espcrc1.tex,v 1.2 2004/02/24 11:22:11 spepping Exp $
\usepackage{graphicx}
\usepackage[figuresright]{rotating}

\newcommand{\AmS}{{\protect\the\textfont2
  A\kern-.1667em\lower.5ex\hbox{M}\kern-.125emS}}

\title{Thermalization through parton transport}

\author{Bin Zhang\address{Department of Chemistry and Physics,
        Arkansas State University, \\
        P.O. Box 419, State University, AR 72467-0419, USA}
        \thanks{This work is supported by the U.S. National
                Science Foundation under Grant No. 0554930.
                We thank the National Energy Research Scientific Computing Center for providing computing resources.
        }}
       
\runtitle{Thermalization through parton transport}
\runauthor{B. Zhang}

\begin{document}

\maketitle

\begin{abstract}
A radiative transport model is used to study kinetic equilibration
during the early stage of a relativistic heavy ion collision. 
The parton system is found to be able to overcome expansion
and move toward thermalization via parton collisions. Scaling
behaviors show up in both the pressure anisotropy and the energy
density evolutions. In particular, the pressure anisotropy
evolution shows an approximate $\alpha_s$ scaling when radiative
processes are included. It approaches an asymptotic time evolution
on a time scale of $1$ to $2$ fm/$c$. The energy density evolution 
shows an asymptotic time evolution that decreases slower than 
the ideal hydro evolution. These observations indicate that 
partial thermalization can be achieved and viscosity is 
important for the evolution during the early longitudinal 
expansion stage of a relativistic heavy ion collision.
\end{abstract}

\section{INTRODUCTION}
Ideal hydrodynamics is very successful in describing experimental
data at the Relativistic Heavy Ion Collider. Ideal
hydrodynamics requires local thermal equilibrium. The ideal
hydrodynamics equations are valid for systems with at least
local isotropy. It is very interesting to study the equilibration
process to get a more detailed understanding of when and how
much thermal equilibration can be achieved in relativistic
heavy ion collisions \cite{Xu:2007aa,Zhang:2008zzk,Huovinen:2008te,Zhang:2008zzu,Zhang:2009rk,El:2009vj}. The isotropization process has been
studied with the AMPT model \cite{Zhang:2008zzk}. 
In particular, the pressure
anisotropy, i.e., the longitudinal to transverse pressure
ratio, is used to measure isotropization in the central
cell in central heavy ion collisions. The string melting 
model is seen to give faster isotropization than 
the default model due to more partons in the initial stage. 
As the parton-parton cross section increases,
isotropization increases. The pressure anisotropy crosses
one at late times when transverse expansion sets in. 

The current AMPT has fixed partonic cross sections and only two 
to two partonic processes. In order to go beyond the current
AMPT description, we introduce the medium dependent
cross sections and particle number changing processes.
The $gg\rightarrow gg$ ($2\rightarrow 2$) 
cross section is taken to be 
the pQCD cross section regulated by a Debye-screening
mass. It is inversely proportional to the screening
mass squared. The screening mass depends on the medium
locally and is related to the density of particles.
As the density increases, the screening mass increases,
and the cross section decreases. 
The $gg\rightarrow ggg$ ($2\rightarrow 3$) cross section
is taken to be one-half of the $2\rightarrow 2$ cross
section, consistent with a more sophisticated
calculation from Xu and Greiner. The $3\rightarrow 2$
reaction integral is determined from detailed balance
to ensure correct chemical equilibration. Isotropic
cross sections will be used and should be reasonable
for the dense central region of a collision.

\begin{figure}[htb]
\begin{minipage}[t]{78mm}
\hspace*{-12mm}
\centering
\includegraphics[scale=0.8]{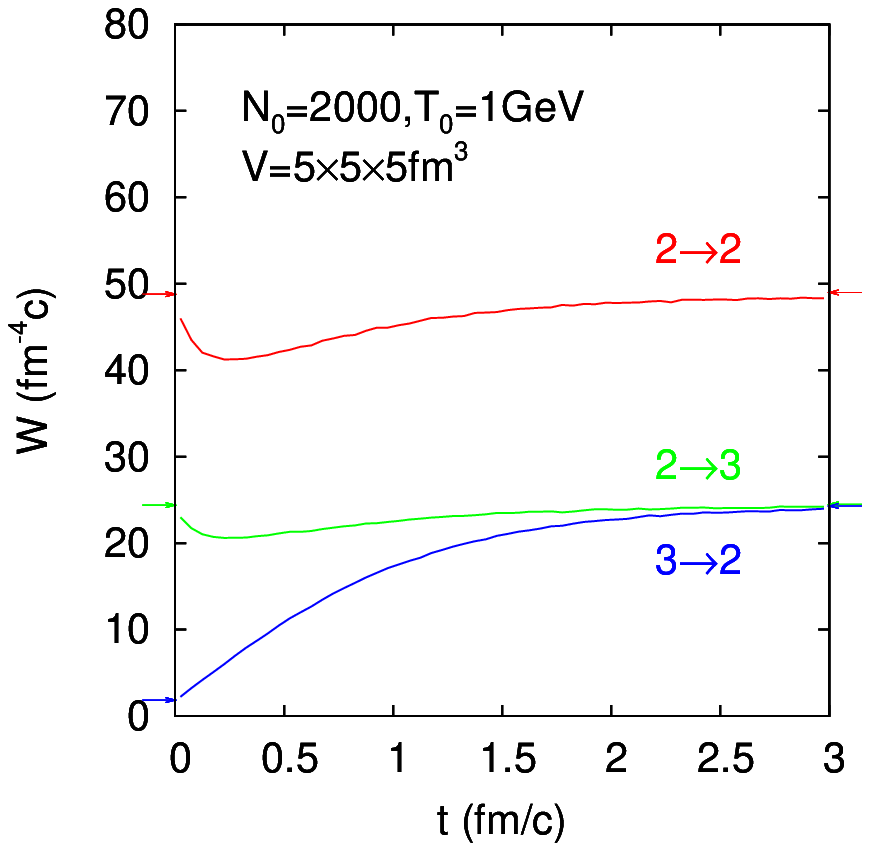}
\caption{Evolutions of collision rates per unit volume. Arrows
indicate initial and chemical equilibrium rates.}
\label{fig:box_rate}
\end{minipage}
\hspace{\fill}
\begin{minipage}[t]{78mm}
\hspace*{-12mm}
\centering
\includegraphics[scale=0.8]{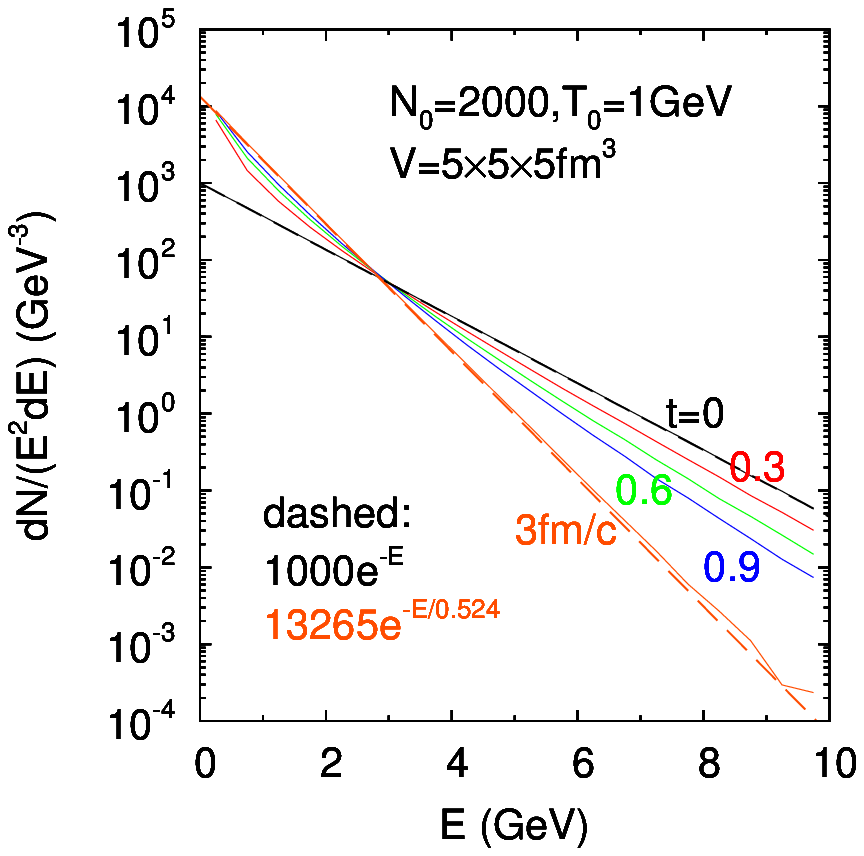}
\caption{Energy distributions at different times. Dashed lines
are initial and chemical equilibrium distributions.}
\label{fig:box_ene}
\end{minipage}
\end{figure}

With this radiative transport, chemical and kinetic
equilibrations can be studied. 
The evolutions of collision rates are shown in 
Fig.~\ref{fig:box_rate} beginning with $2000$
particles in a box of $5\times 5\times 5$ fm$^3$ with
a temperature of $1$ GeV and coupling constant
$\alpha_s=0.4$. The
initial $2\rightarrow 3$ rate is large than 
$3\rightarrow 2$ rate and there will be particle 
production. As time goes on, these two rates converge
toward the rate in chemical equilibrium. It is
interesting to see whether kinetic equilibrium is
maintained throughout the chemical equilibration
process. Fig.~\ref{fig:box_ene} shows that the 
energy distribution deviates from an exponential 
distribution at early times. In particular, at $0.3$
fm/$c$, low energy part is closer to the final 
equilibrium value. The system relaxes toward the
final thermodynamics equilibrium and at time
$3$ fm/$c$, it is very close to the expected 
distribution with chemical and kinetic equilibrium.

\section{RESULTS}

\begin{figure}[htb]
\begin{minipage}[t]{80mm}
\hspace*{-12mm}
\centering
\includegraphics[scale=0.8]{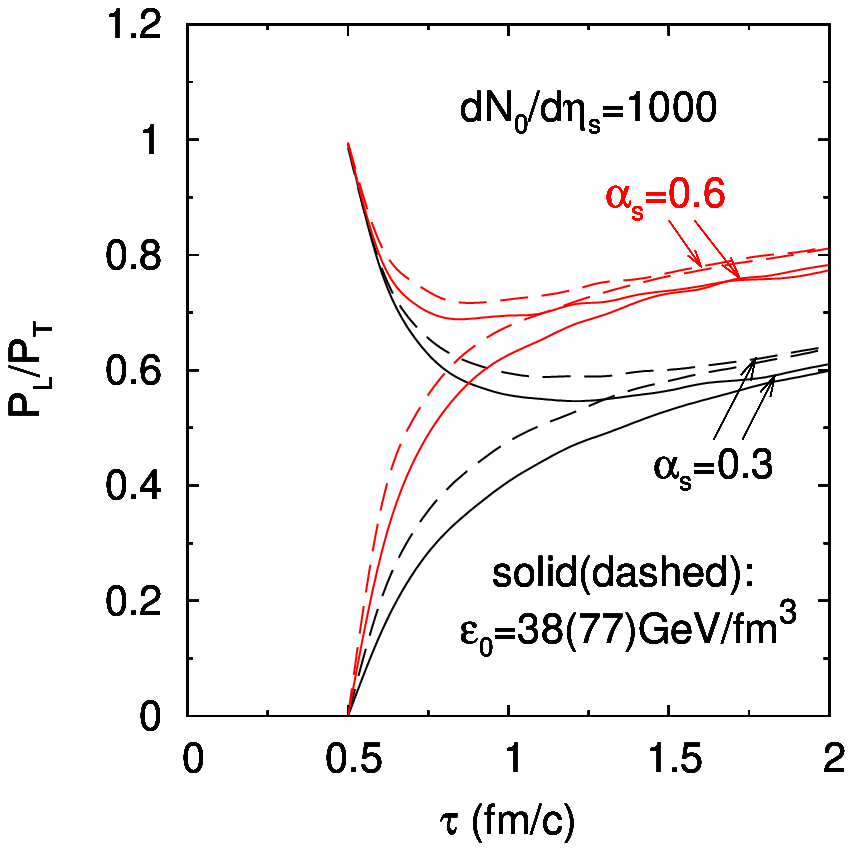}
\caption{Press anisotropy evolution.}
\label{fig:bj_plopt}
\end{minipage}
\hspace{\fill}
\begin{minipage}[t]{75mm}
\hspace*{-12mm}
\centering
\includegraphics[scale=0.8]{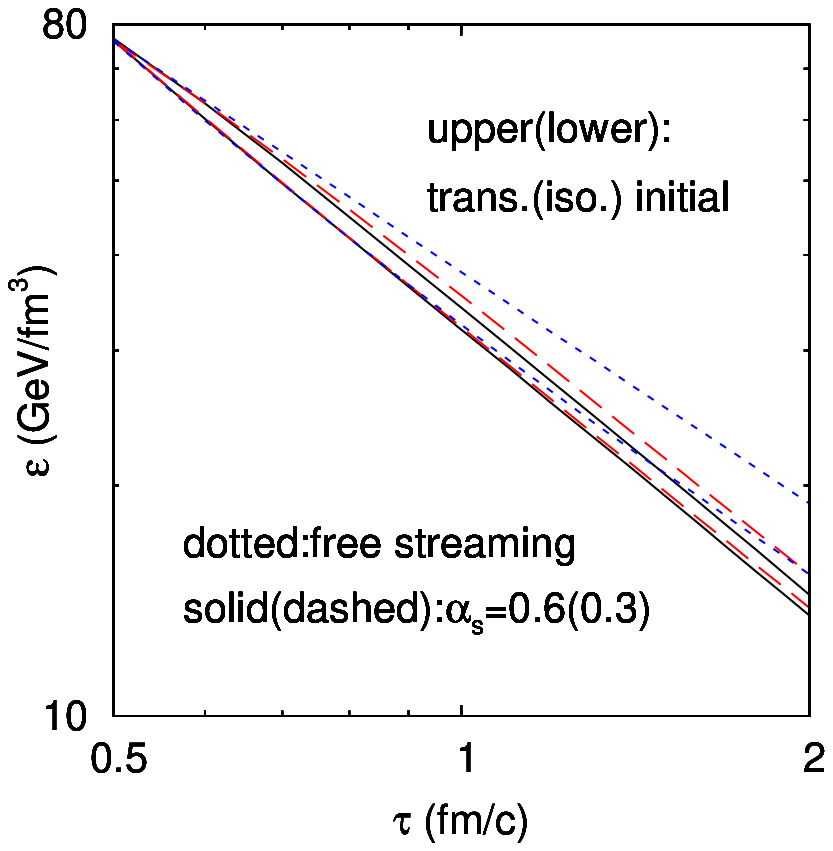}
\caption{Energy density evolution.}
\label{fig:bj_ene}
\end{minipage}
\end{figure}

Kinetic equilibration during the early stage of a heavy ion
collision can be studied with the radiative transport
model \cite{Zhang:2009rk}. We will start with either
initial isotropic or initial transverse distribution. The initial
particle distributions are all exponential. Two initial energy
densities $\epsilon_0=38,77$ GeV/fm$^3$ will be used. The former
corresponds to an initial temperature of $T_0=0.5$ GeV for the 
isotropic initial conditions, while the latter corresponds to
$1$ GeV. Two values are used for the strong interaction coupling
constant $\alpha_s=0.3,0.6$. Fig.~\ref{fig:bj_plopt} shows the
pressure anisotropy ratio evolutions. With isotropic initial 
conditions,
$P_L/P_T$ first decreases due to the influence of longitudinal
expansion then increases as thermalization wins over expansion.
In contrast to the case with only $2\rightarrow 2$ collisions,
instead of the $\alpha_sT_0$ scaling, there is an approximate
$\alpha_s$ scaling. Furthermore, the transverse initial 
condition case has the same asymptotic evolution as the
isotropic initial condition case. In other words, the memory
of the initial pressure anisotropy is lost after some 
relaxation time. As the initial energy density increases or
as the coupling constant increases, the pressure anisotropy
goes closer to one and there is more thermalization and more
memory loss of initial conditions. The energy density evolutions
are shown in Fig.~\ref{fig:bj_ene}. This is for the high initial
energy density case, similar results are obtained for the low
initial energy density case. Fig.~\ref{fig:bj_ene} shows that 
the energy density evolution at very early times is determined
by the initial pressure anisotropy and the late time evolution
is determined by final state interactions. In particular,
we observe that the evolutions are bounded by $1/\tau$
for the free streaming case and $1/\tau^{1.25}$ when 
$\alpha_s=0.6$. This falls
a little short of the ideal hydro $1/\tau^{1.33}$ behavior
and shows that viscosity is important in the longitudinal
expansion stage.

\end{document}